\documentclass[seceq,twocolumn]{jpsj2}
\usepackage{epsf}

\def\runtitle{Probability-Changing Cluster Algorithm: 
3D Ising Model and Percolation Problem}
\def\runauthor{Yusuke {\sc Tomita} and Yutaka {\sc Okabe}}

\title{Probability-Changing Cluster Algorithm: 
Study of Three-Dimensional \\
Ising Model and Percolation Problem}

\author{Yusuke {\sc Tomita}\footnote{E-mail: 
ytomita@phys.metro-u.ac.jp} 
and Yutaka {\sc Okabe}\footnote{E-mail: okabe@phys.metro-u.ac.jp}}

\inst{
Department of Physics, Tokyo Metropolitan University,
Hachioji, Tokyo 192-0397
}

\recdate{January 7, 2002}

\abst{
We present a detailed description of the idea and procedure for 
the newly proposed Monte Carlo algorithm of tuning the critical 
point automatically, which is called 
the probability-changing cluster (PCC) algorithm 
[Y. Tomita and Y. Okabe, Phys. Rev. Lett. {\bf 86} (2001) 572]. 
Using the PCC algorithm, we investigate the three-dimensional 
Ising model and the bond percolation problem.  
We employ a refined finite-size scaling analysis to make 
estimates of critical point and exponents.  With much less efforts, 
we obtain the results which are consistent with the previous calculations. 
We argue several directions for the application of 
the PCC algorithm. 
}

\kword{Monte Carlo simulation, cluster algorithm, Ising model, 
percolation, finite-size scaling
}

\begin{document}
\sloppy
\maketitle

\section{Introduction}
The Ising model \cite{Ising} is a basic model for studying 
phase transitions and critical phenomena. 
The $q$-state Potts model \cite{Potts,Wu}, 
which has $q$ components for the order parameter, 
is a generalization of the Ising model. 
Then, the Ising model corresponds to the $q=2$ Potts model. 
The percolation model \cite{Stauffer} also exhibits 
a geometric phase transition.  
The Kasteleyn-Fortuin (KF) cluster representation \cite{KF}
of the $q$-state Potts model bridges the Potts model 
and the percolation problem; the bond percolation problem 
can be regarded as the $q=1$ Potts model. 
Recently, based on the cluster formalism, the multiple-percolating 
clusters of the Ising system with large aspect ratio have been 
studied \cite{TOH}.

For two-dimensional (2D) systems, exact or rigorous results have been 
obtained for the Potts models \cite{Onsager,Beale}, and they are used 
as the testing ground for numerical study. 
On the other hand, for three-dimensional (3D) systems, 
it is rare that exact results are available, 
and we rely more on numerical studies for revealing the nature 
of the problem.
The Monte Carlo simulation \cite{LanBin} is a standard powerful tool 
to study critical phenomena numerically. 
To obtain accurate data, the development of efficient algorithms is 
highly demanded.
Cluster algorithms \cite{SW,Wolff} are examples of such efforts, 
and they have been successfully used to overcome
slow dynamics in the Monte Carlo simulation.
Swendsen and Wang (SW) \cite{SW} applied the KF \cite{KF}
representation to identify clusters of spins.

Extending the SW algorithm, we have recently proposed 
an effective cluster algorithm of tuning the critical point 
automatically; this algorithm is called 
the probability-changing cluster (PCC) algorithm \cite{PCC}. 
We have shown the effectiveness of the PCC algorithm for 
the case of 2D Potts models in the Letter \cite{PCC}. 
The basic idea of our algorithm is that we change the probability
of connecting parallel spins $p$ in the KF representation during 
the process of the Monte Carlo spin update. 
We decrease or increase $p$ depending on the observation whether
the KF clusters are percolating or not percolating; 
essentially, we change the temperature. 
This simple negative feedback mechanism together with the
finite-size scaling (FSS) \cite{Fisher} property of 
the existence probability (also called the crossing probability) $E_p$, 
the probability that the system percolates, 
leads to the determination of the critical point.
Since our ensemble is asymptotically canonical as $\Delta p$,
the amount of the change of $p$, becomes $0$, the distribution
functions of physical quantities obey the FSS; as a result,
we can determine critical exponents using the FSS analysis.

Previously, Machta {\it et al.} \cite{machta95} proposed another 
idea of cluster algorithm to tune the critical point automatically, 
which is called the invaded cluster (IC) algorithm. 
However, the ensemble of the IC algorithm is not necessarily clear, 
and it has a problem of ``bottlenecks", which causes the broad tail 
in the distribution of the fraction of 
the accepted satisfied bonds \cite{machta95}. 
In contrast, it is guaranteed that we approach the canonical 
ensemble in our PCC algorithm. 

In this paper we give the more detailed description of 
the PCC algorithm, and show the results for the 3D Ising model 
and the 3D bond percolation problem.  We pay attention to 
the refined FSS analysis for determining the critical point 
and critical exponents, which is the same idea 
as was used in a high-resolution 
Monte Carlo study by Ferrenberg and Landau \cite{FerLan}. 
The rest of this paper is organized as follows: 
In \S 2 we explain the idea and the procedure of the PCC algorithm.
In \S 3 the results for the 3D Ising model and percolation problem 
are shown, and the refined FSS analysis is discussed.
Finally, we summarize this paper and give discussions in \S 4.

\section{Probability-Changing Cluster Algorithm}
\subsection{Idea of PCC algorithm for Ising model}
We start with explaining the idea of the PCC algorithm.
Here, we deal with the ferromagnetic Ising model, 
whose Hamiltonian is given by
\begin{equation}
  {\cal H} = -J\sum_{\left<i,j\right>} \sigma_i\sigma_j,
  \quad \sigma_i=\pm1,
\end{equation}
where $J$ is the exchange coupling constant, and the summation 
is taken over the nearest-neighbor pairs $\left<i,j\right>$. 
In this case, the probability of connecting parallel spins 
in the KF representation is given by $p = 1-e^{-2J/k_BT}$. 
The procedure of Monte Carlo spin update is as follows: 
\begin{enumerate}
\item[1.]  Start from some spin configuration and some value of $p$. 
\item[2.]  Make KF clusters using the probability $p$,
and check whether the system is percolating
or not. Update spins following the same rule as the SW algorithm,
that is, flip all the spins on any KF cluster to one of two states.
\item[3.]  If the system is percolating (not percolating) 
in the previous test,
decrease (increase) $p$ by $\Delta p$ $(>0)$. 
Essentially, we change the temperature $T$. 
\item[4.]  Go back to the process 2.
\end{enumerate}
Since we use the cluster representation and assign clusters, 
we are ready to check whether the system is percolating or not 
in the process 2.  
The distribution of $p$ for Monte Carlo samples approaches 
the Gaussian distribution of which mean value is $p_c(L)$, 
after repeating the above processes.  
Here, $p_c(L)$ is the probability of connecting spins,
such that the existence probability $E_p$ becomes $1/2$, 
and depends on the system size $L$. 
The existence probability $E_p$ is the probability that the system percolates.
In the limit of $\Delta p \to 0$, we approach the canonical ensemble, 
which will be discussed later. 
Then, we can use the FSS analysis.
We should note that $E_p$ follows the FSS near the critical point,
\begin{equation}
 E_p(p,L) \sim X(tL^{1/\nu}),
 \quad t=(p_c-p)/p_c,
\label{existence}
\end{equation}
as far as the corrections to FSS are negligible, 
where $p_c$ is the critical value of $p$ for the infinite system
$(L \to \infty)$ and $\nu$ is the correlation-length critical exponent. 
Then, we can estimate $p_c$ from the size dependence of $p_c(L)$ using
eq.~(\ref{existence}) and, in turn, estimate $T_c$ through the relation
$p_c=1-e^{-2J/k_BT_c}$.

\subsection{Percolating condition}
There are several choices of criterion to determine percolating.  
For example, Machta {\it et al.} \cite{machta95} used both the extension 
rule and the topological rule for their stopping condition 
in the IC algorithm.
The former rule is that some cluster has maximum extent $L$ 
in at least one of the $d$ directions in $d$-dimensional systems.
The latter rule is that some cluster winds around the system 
in at least one of the $d$ directions.  We may use 
these percolating conditions in our PCC algorithm.  
Actually, we can use any rule 
to determine percolating, but FSS functions for physical quantities, 
therefore $p_c(L)$, depend on the rule.  

\subsection{Distribution of $p$}

Let us consider the distribution of $p$, $f(p)$. 
We change $p$ based on the observation whether the system 
is percolating or not, which leads to the negative feedback 
mechanism.  Since the existence probability 
$E_p(p)$ is the probability that the systems percolates, 
the transition probabilities $W$ from $p$ to $p+\Delta p$ 
and from $p$ to $p-\Delta p$ in the PCC algorithm 
are written as follows: 
\begin{equation}
 \left\{ 
   \begin{array}{l}
     W_{p \to p+\Delta p} = 1 - E_p(p), \vspace{2mm} \\
     W_{p \to p-\Delta p} = E_p(p).
   \end{array}
 \right.
\label{trans}
\end{equation}
In the vicinity of $p_c(L)$, we may employ the linear 
approximation for $E_p(p)$, such as
\begin{equation}
 E_p(p) = \frac{1}{2} + a(p-p_c(L)),
\label{Ep_linear}
\end{equation}
where $a$ is the value of $dE_p/dp$ at $p_c(L)$. 
Then, this problem is nothing but the Ehrenfest model for 
{\it diffusion with a central force} \cite{Ehrenfest,Feller}. 
In the steady state, the distribution of $p$, $f(p)$, satisfies 
the relation
\begin{equation}
 f(p) = W_{p-\Delta p \to p} f(p-\Delta p) +
        W_{p+\Delta p \to p} f(p+\Delta p). 
\label{fp}
\end{equation}
In the linear approximation, eq.~(\ref{Ep_linear}), 
$p$ takes the values between $p_c(L)-1/2a$ 
and $p_c(L)+1/2a$.
Substituting eqs.~(\ref{trans}) and (\ref{Ep_linear}) into 
eq.~(\ref{fp}) and denoting $p=p_c(L) + i \Delta p$, 
we can show that $f(p)$ is the binomial distribution, that is, 
\begin{equation}
 f(p) \propto \; _n C_{n/2+i},
\end{equation}
where $n=(1/a)/\Delta p$, $i \in [-n/2,n/2]$, 
and $_n C_{n/2+i}$ are the binomial coefficients. 
Thus, for large $n$, or small $\Delta p$, the distribution function $f(p)$ 
becomes the Gaussian distribution with the average $p_c(L)$ 
and the variance $\sigma^2 = (n/4) \, (\Delta p)^2 = 
\Delta p/4a$. 
For smaller $\Delta p$, the width of the distribution becomes 
narrower as $\sigma \propto \sqrt{\Delta p}$. 
Since $a$ is the value of $dE_p/dp$ at $p_c(L)$, 
we expect $a \propto L^{1/\nu}$ 
using the FSS assumption, eq.~(\ref{existence}). 
Thus, for larger $L$, the width of $f(p)$ becomes narrower. 

In the Letter \cite{PCC}, we have checked that the distribution 
of $p$ actually approaches the Gaussian distribution 
with very narrow width for the 2D Ising model, and the resulting 
energy histogram is indistinguishable from that 
by the constant-temperature calculation. 
Moreover, we have obtained the expected 
$\Delta p$- and $L$-dependence for the width of $f(p)$. 

\subsection{Determination of next $p$}
In the process 3, we decrease or increase $p$ by $\Delta p$. 
The difference $\Delta p$ is a free parameter in our algorithm. 
In the limit of small $\Delta p$ we approach the canonical 
ensemble, but it takes a long time to equilibrate for small $\Delta p$.
Using the same approximation as in the previous subsection and
assuming that the subsequent steps are independent, 
we can show that the deviation of the average value of $p$ 
from $p_c(L)$ becomes smaller as a geometric progression with time.  
Since the geometric ratio is given by $1-2a\Delta p$ 
in this approximation, 
the convergence becomes slower for smaller $\Delta p$.  
Practically, we may start with rather large $\Delta p$, and gradually
decrease $\Delta p$ with monitoring the trail of the values of $p$.
Small steps of preparation are enough for equilibration.

We change $p$ by every Monte Carlo step in our original 
proposal \cite{PCC}.  As another way, we may measure 
the existence probability $E_p$ for a short time interval 
with keeping $p$ constant, and then change $p$ 
for the next short time interval.  
In this process, recording the values of $p$ and whether percolation 
occurred for each, we may extract the information on the final 
$p_c(L)$ efficiently using the Bayesian statistics \cite{Swendsen}.  
A more deterministic way of adjusting $p_c(L)$ may be considered.  
Solving $E_p(p)=1/2$ iteratively by the Newton method 
may lead to the adjustment of $p_c(L)$ \cite{Wang}. 
It is quite interesting to improve and extend the method of 
PCC algorithm. 

\subsection{Checking value of $E_p$}

We have chosen the value of $E_p$ which gives $p_c(L)$ as 1/2 
because it is the simplest.
In case the critical value of $E_p$ at the critical point of the 
infinite system is far from 1/2, it is convenient to employ 
the checking value of $E_p$ different from 1/2. 
We may modify the update process such that this value is 
different from 1/2. If we want to choose the checking value of $E_p$ 
as $e_p$, we may modify the process 3 as follows: 
\begin{enumerate}
\item[3'.] 
If the system is percolating in the previous test, 
decrease $p$ by $\Delta p$ with the probability $s=\min[(1/2)/e_p,1]$ 
and increase $p$ with the probability $1-s$. 
On the contrary, if the system is not percolating, 
increase $p$ by $\Delta p$ with the probability $s'=\min[(1/2)/(1-e_p),1]$ 
and decrease $p$ with the probability $1-s'$.
\end{enumerate}
In this way, we can control the checking value of $E_p$.

\section{Results} 
\subsection{3D Ising model}
Here we present the result of the 3D Ising model.  We have simulated 
the Ising model on the simple cubic lattice by using the PCC algorithm. 
We have treated the systems with linear sizes $L$=8, 12, 16, 24, 32, 48,
and 64.  For the criterion to determine percolating, we have employed 
the topological rule \cite{machta95} in the present study.
After 10,000 Monte Carlo sweeps of determining $p_c(L)$ with
gradually reducing $\Delta p$, we have made 100,000 Monte Carlo sweeps
to take the thermal average; we have made 10 runs for each size
to get better statistics and to evaluate the statistical errors.
We have started with $\Delta p = 10^{-3}$ for $L=8$ and 
$\Delta p = 10^{-4}$ for $L=64$. 
For the intermediate sizes, we have started with $\Delta p$ 
between these two values. 
The final value of $\Delta p$ has been chosen as 
$\Delta p = 10^{-5} \times L^{-1.6}$ for the system size $L$. 
Actually, the schedule of decreasing $\Delta p$ is not so serious. 

\begin{figure}[tb]
\epsfxsize=8.5cm
\centerline{\epsfbox{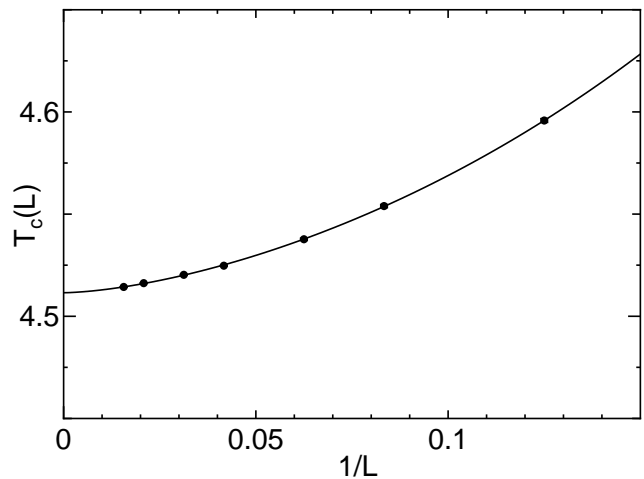}}
\caption{Plot of $T_c(L)$ (in units of $J/k_B$) as a function of $1/L$
for the 3D Ising model. The system sizes are $L=8, 12, 16, 24, 32, 48$,
and $64$.}
\label{fig_1}
\end{figure}

We plot the size-dependent $T_c(L)$ as a function of $1/L$ for the 3D
Ising model in Fig.~\ref{fig_1}. From now on, we represent the temperature in units
of $J/k_B$. The error bars are within the size of the mark. 
The critical temperature $T_c$ can be estimated by the FSS relation, 
eq.~(\ref{existence}).  Including the corrections to FSS, 
we have 
\begin{equation}
 T_c(L) = T_c + aL^{-1/\nu}(1 + bL^{-\omega}), 
\label{tc}
\end{equation}
where $T_c(L)$ is given through $p_c(L)=1-e^{-2J/k_BT_c(L)}$, and 
$\omega$ is the correction-to-FSS exponent.  
Since there are five fitting parameters in eq.~(\ref{tc}), it is not easy
to get accurate estimates of the critical point and critical exponents.
In a high-resolution Monte Carlo study, Ferrenberg and Landau \cite{FerLan}
employed a FSS analysis to get accurate estimates. 
They first determine the exponent $\nu$, and with $\nu$ determined 
quite accurately they then estimate $T_c$. 
Since their procedure is well fitted for our algorithm, 
we use the same idea.  

We first note that the inverse-temperature derivatives of the logarithm 
for the moment of magnetization $m$ obey the following FSS relations,
\begin{equation}
 \frac{\partial \ln\left<m^n\right>}{\partial K} =
 a^{\prime}L^{1/\nu} (1 + b^{\prime}L^{-\omega}),
\label{logder}
\end{equation}
where $K=1/k_BT$ and $\left<\cdots\right>$ denotes the thermal average. 
These FSS relations hold for the values of $m$ at the size-dependent 
$T_c(L)$ as well as those at the fixed $T_c$ for the infinite system.  
Here, we treat the variables at $T_c(L)$. 
Since we calculate the left-hand side of eq.~(\ref{logder}) by 
the general formula to calculate the $K$-derivative of any 
quantity $\left< A \right>$,
\begin{equation}
 \frac{\partial \left<A \right>}{\partial K} =
 \left<A\right>\left<E\right> - \left<AE\right>,
\label{der-K}
\end{equation}
where $E$ is the energy, we can extract $\nu$ 
without determining $T_c$ for the infinite system. 
The inverse-temperature derivative of the Binder parameter \cite{Binder}, 
which is defined as
\begin{equation}
 g = \frac{1}{2}\left(3-\frac{\left<m^4\right>}{\left<m^2\right>^2}\right),
\label{Binder}
\end{equation}
also obeys the FSS relation as in eq.~(\ref{logder}).  
What we do is that we measure the inverse-temperature derivatives 
of $\ln \left<m^n\right>$ and $g$ at the size-dependent 
$T_c(L)$ and make an analysis based on the FSS relations, 
eq.~(\ref{logder}).

\begin{figure}[tb]
\epsfxsize=8cm
\centerline{\epsfbox{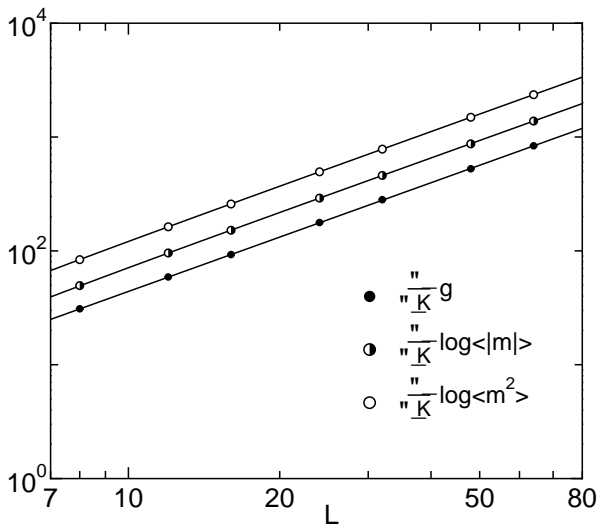}}
\caption{Plot of derivatives at $T_c(L)$ as a function of $L$ for
the 3D Ising model in logarithmic scale.}
\label{fig_2}
\end{figure}

We plot the derivatives $\partial g/\partial K$, 
$\partial \ln \left<|m| \right>/\partial K$ and 
$\partial \ln \left<m^2 \right>/\partial K$ at $T_c(L)$ 
as a function of $L$ in logarithmic scale in Fig.~\ref{fig_2}.
The error bars are again within the size of the mark.
We find the power-law size dependence from the linearity of the data 
with small corrections to FSS; 
we estimate the exponent $1/\nu=1.594(8)$ using eq.~(\ref{logder}).
Here, the number in the parenthesis denotes the uncertainty in the
last digit.  We have used the average of three data.  
The correction-to-FSS exponent $\omega$ is 1.2(5), 
which is a little bit larger than the recent value 
$\omega = 0.87(9)$ \cite{Balles}.
Our estimate of $T_c$ using eq.~(\ref{tc}) is 
$T_c=4.5108(7)$, that is, $1/T_c=0.22169(4)$.
Both estimates of $T_c$ and $\nu$ are consistent with the estimates
of the recent study \cite{FerLan},
$1/\nu=1.590(2)$ and $1/T_c=0.2216595(26)$. 
The solid curve in Fig.~\ref{fig_1} is the best fitted curve 
for eq.~(\ref{tc}) with $\nu$ determined accurately first.

\begin{figure}[tb]
\epsfxsize=8cm
\centerline{\epsfbox{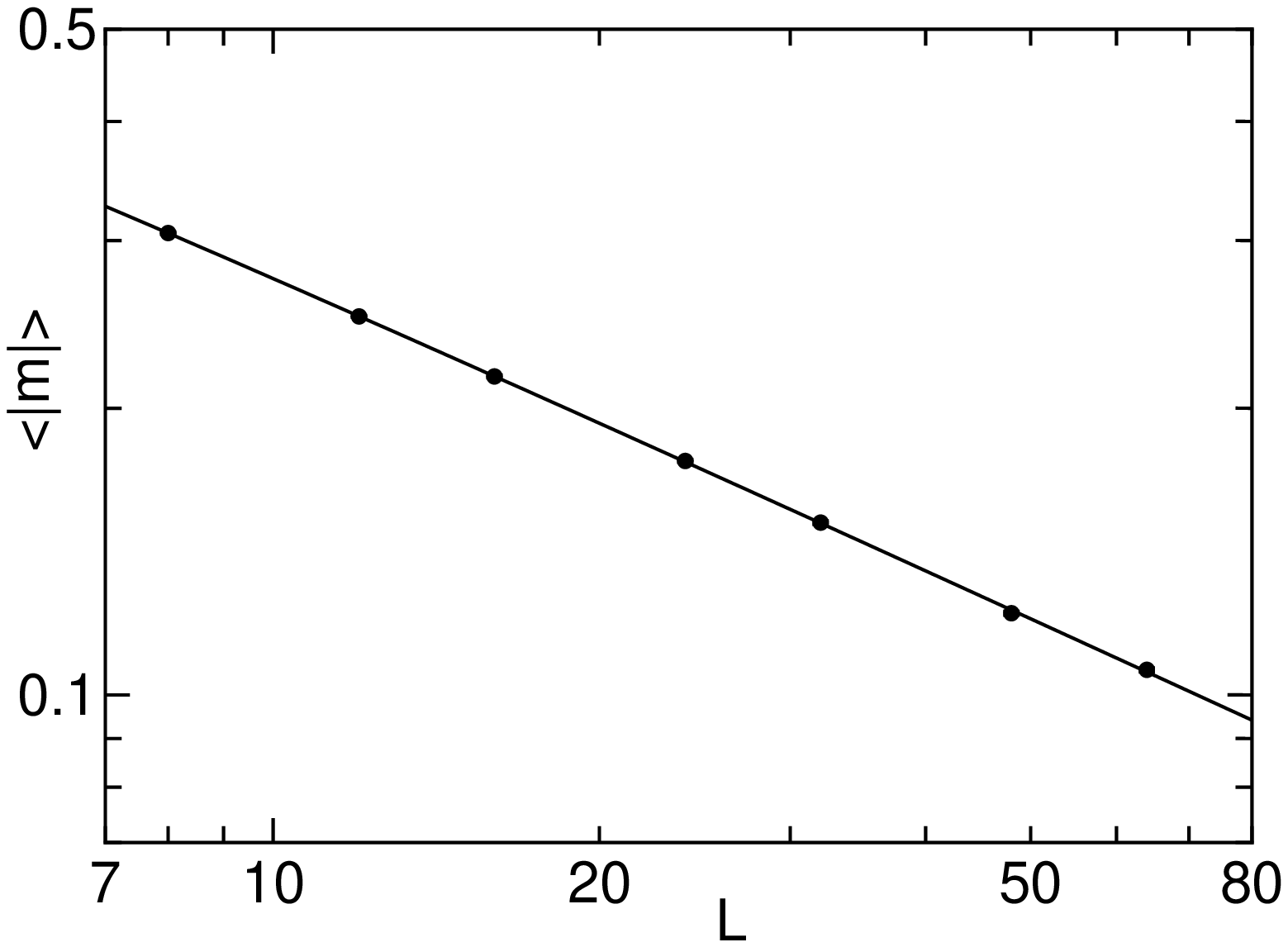}}
\caption{Plot of $\left<|m|\right>$ at $T_c(L)$ as a function of $L$ for
the 3D Ising model in logarithmic scale.}
\label{fig_3}
\end{figure}

In order to discuss another exponent, we plot 
the average of the magnetization $\left<|m|\right>$ at $T_c(L)$
as a function of $L$ in logarithmic scale in Fig.~\ref{fig_3}. 
We use the FSS relation with the corrections to FSS,
\begin{equation}
 \left<|m|\right>_{T=T_c(L)} = cL^{-\beta/\nu}(1 + dL^{-\omega}),
\label{mag}
\end{equation}
for the estimate of the magnetization exponent 
$\beta$.  From the least square fit using eq.~(\ref{mag}), 
we have $\beta/\nu=0.517(8)$, which is again consistent with 
the recent estimate \cite{FerLan}, $\beta/\nu=0.518(7)$.

\begin{figure}[tb]
\epsfxsize=8cm
\centerline{\epsfbox{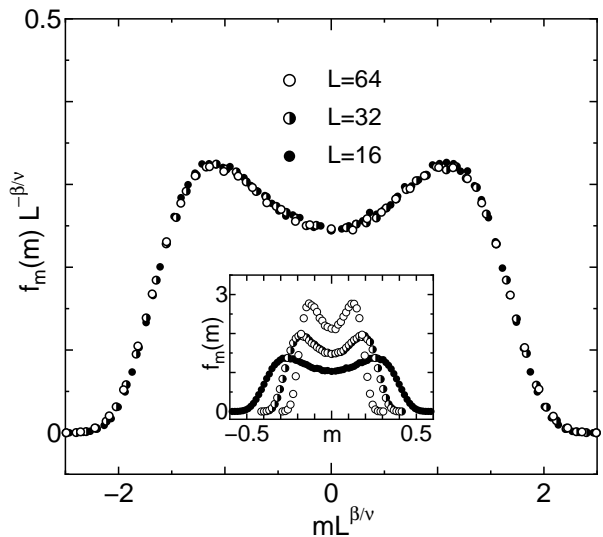}}
\caption{FSS plot of $f_m(m)$ for the 3D Ising model, 
where $\beta/\nu=0.517$. The system sizes are $L=16, 32$, and $64$.
The inset shows the raw data.}
\label{fig_4}
\end{figure}

It is interesting to study the distribution function of physical quantities.
We show the FSS plot of the distribution function $f_m(m)$ in
Fig.~\ref{fig_4}, based on the FSS relation,
\begin{equation}
 f_m(m)_{T=T_c(L)} \sim L^{\beta/\nu}h(mL^{\beta/\nu}).
\end{equation}
The inset of Fig.~\ref{fig_4} shows the raw data 
of the distribution functions $f_m(m)$ 
for linear sizes $L=16, 32$, and $64$. 
The scaled data show very good FSS behavior; that is, the data 
of different sizes are collapsed on a single curve.

\subsection{3D bond percolation problem}
The idea of the PCC algorithm is based only on the property of
a percolation problem. Thus, it is straightforward, or even easier,
to apply this algorithm to the geometric percolation problem. 
The partition function for the bond percolation problem is 
written as 
\begin{equation}
 Z = \sum_{G^{\prime}\subseteq G}p^{b(G^{\prime})} (1-p)^{N_b-b(G^{\prime})}.
\label{Z}
\end{equation}
Here $G$ is all the configurations, or the graph, and ${b(G^{\prime})}$ is 
the number of occupied bonds in the subgraph $G^{\prime}$. 
The sum is over all subgraphs $G^{\prime}$ of $G$.
The probability of bond occupation is denoted by $p$, and 
$N_b$ is the total number of bonds in the system. 
In the bond percolation problem, we are to locate 
the percolation threshold $p_c$. 
We change $p$ by the small amount of $\Delta p$ in the process 
of simulation, and determine the size-dependent $p_c(L)$ 
automatically as in the case of the Ising model.  
In the limit of $\Delta p \to 0$, it becomes 
the usual percolation problem.
One thing we should have in mind is that we determine whether 
the bond is occupied or not with the probability $p$ for each bond.  

\begin{figure}[tb]
\epsfxsize=8cm
\centerline{\epsfbox{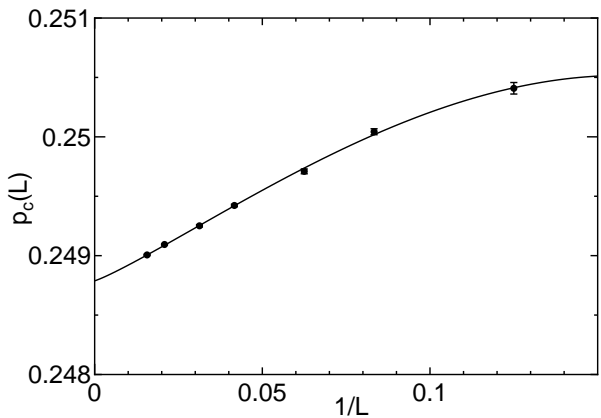}}
\caption{Plot of $p_c(L)$ as a function of $1/L$
for the 3D bond percolation problem. 
The system sizes are $L=8, 12, 16, 24, 32, 48$, and $64$.}
\label{fig_5}
\end{figure}

We have studied the 3D bond-percolation model with linear sizes 
$L=8, 12, 16, 24, 32, 48,$ and $64$.  Almost the same conditions 
are used as in the 3D Ising model. 
We plot the size-dependent $p_c(L)$ as a function of $1/L$ 
in Fig.~\ref{fig_5}.  The FSS relation for $p_c(L)$ is given 
by the equation similar to eq.~(\ref{tc}), but we follow 
the same scheme as in the Ising model to get better estimates 
of the critical point and exponents; that is, 
we first estimate the critical exponent $\nu$.  
The fraction of lattice sites in the largest cluster $c$ plays a role 
of the order parameter.  Thus, we consider the $p$-derivative of 
the moments of $c$.  The derivative of any quantity $\left<A\right>$ 
with respect to $p$ is given by
\begin{equation}
 \frac{\partial \left<A\right>}{\partial p} =
 \frac{\left<Ab\right>-\left<A\right>\left<b\right>}{p(1-p)},
\label{der-p}
\end{equation}
where $b$ is the number of occupied bonds, and $\left< \cdots \right>$ 
denotes the sample average. 
Equation~(\ref{der-p}) corresponds to the general formula for 
the $K$-derivative, eq.~(\ref{der-K}). 
We may derive eq.~(\ref{der-p}) starting from the expression for 
the partition function, eq.~(\ref{Z}), or using the correspondence 
based on the KF relation, $p=1-e^{-2J/k_BT}$.

\begin{figure}[tb]
\epsfxsize=8cm
\centerline{\epsfbox{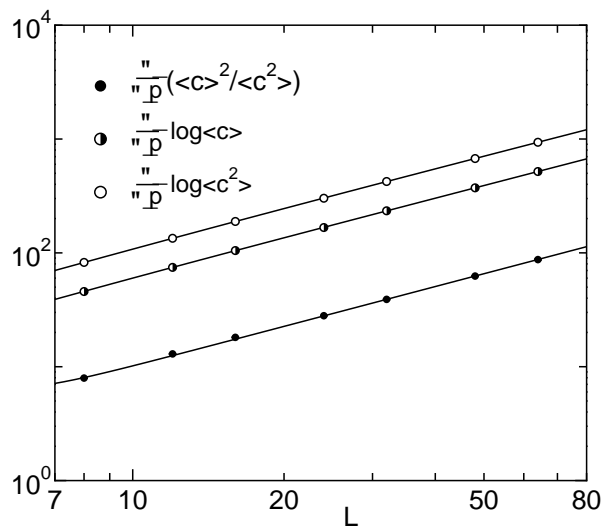}}
\caption{Plot of derivatives at $p_c(L)$ as a function of $L$ for
the 3D bond percolation problem in logarithmic scale.}
\label{fig_6}
\end{figure}

We have calculated the logarithmic derivatives of 
$\left<c\right>$ and $\left<c^2\right>$, where $c$ is the 
fraction of lattice sites in the largest cluster.
Since the moment ratio $\left<c\right>^2/\left<c^2\right>$ 
has the same FSS property as the Binder parameter, eq.~(\ref{Binder}),
we also calculate the $p$-derivative of this moment ratio.
We plot the derivatives 
$\partial (\left<c \right>^2/\left<c^2 \right>)/\partial p$, 
$\partial \ln \left<c \right>/\partial p$ and 
$\partial \ln \left<c^2 \right>/\partial p$ at $p_c(L)$ 
as a function of $L$ in logarithmic
scale in Fig.~\ref{fig_6}. 
We find the power-law size dependence from the linearity of the data 
with small corrections to FSS; 
we estimate the exponent $1/\nu=1.12(5)$ from the slopes of lines.
The correction-to-FSS exponent $\omega$ is 1.1(5), 
which is a little bit smaller than the recent estimate for site percolation 
problem, $\omega=1.62(13)$ \cite{Balles}.
We can use the FSS form similar to eq.~(\ref{tc}) for the estimate of $p_c$;
our estimate using the value of $\nu$ is $p_c=0.24881(3)$.
Both estimates of $p_c$ and $\nu$ are consistent with the estimates
of the recent study \cite{LorZif}, $1/\nu=1.12(3)$ and $p_c=0.2488126(5)$.
The solid curve in Fig.~\ref{fig_5} is the best fitted curve 
for $p_c$, which is similar to eq.~(\ref{tc}), 
with $\nu$ determined accurately first.

\begin{figure}[tb]
\epsfxsize=8cm
\centerline{\epsfbox{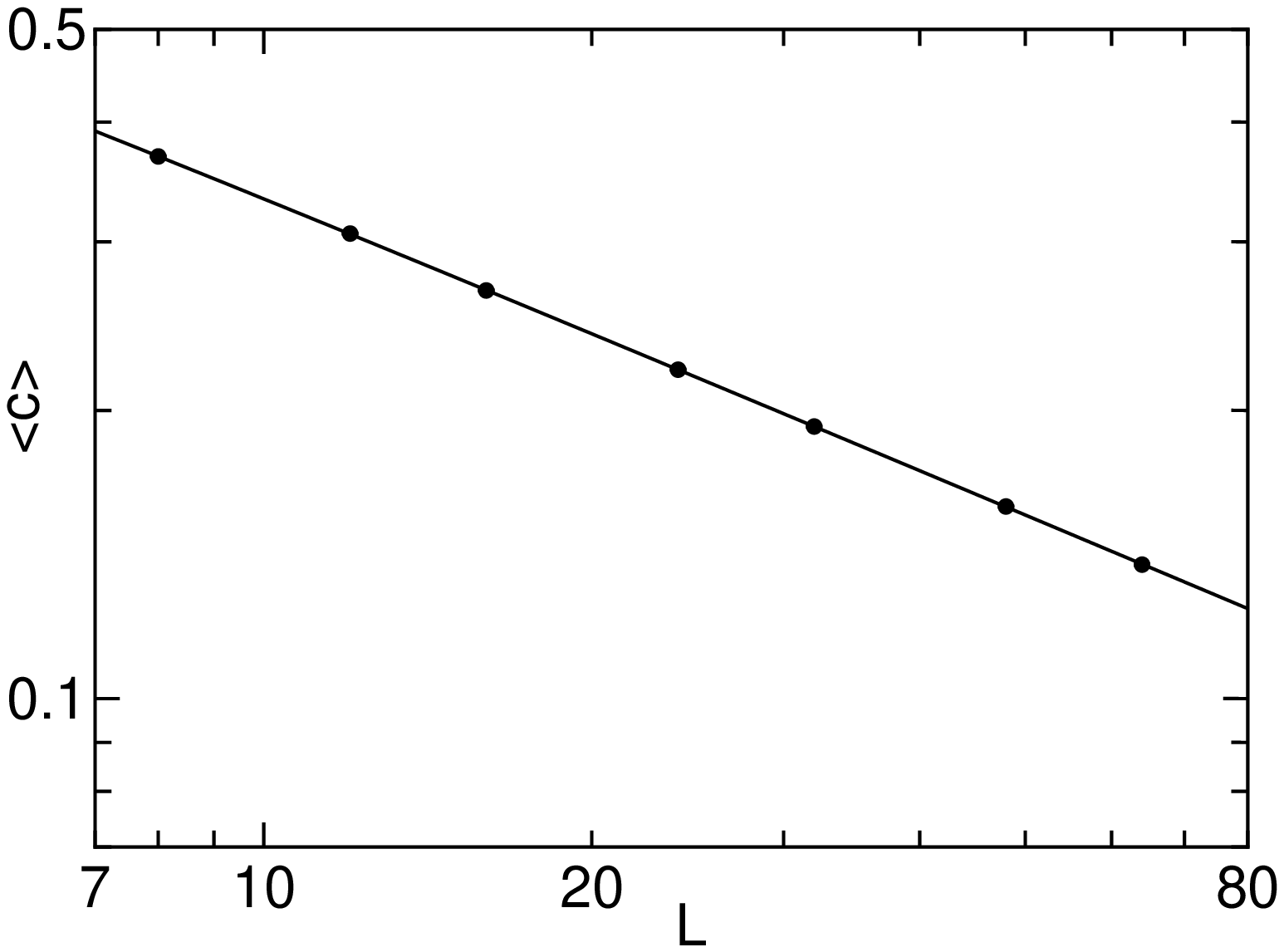}}
\caption{Plot of $\left<c\right>$ at $p_c(L)$ as a function of $L$ for
the 3D bond percolation problem in logarithmic scale.}
\label{fig_7}
\end{figure}

We plot the fraction of lattice sites in the largest cluster
$\left<c\right>$ at $p_c(L)$ as a function of $L$
in logarithmic scale in Fig.~\ref{fig_7}. 
Since $\left<c\right>$ has the same FSS form as the magnetization,
we use eq.~(\ref{mag}) for the estimate of $\beta/\nu$.
Using the least square fit, we have
$\beta/\nu=0.474(5)$, which is again consistent with the estimate of
recent study \cite{LorZif}, $\beta/\nu=0.476(5)$.

\begin{figure}[tb]
\epsfxsize=8cm
\centerline{\epsfbox{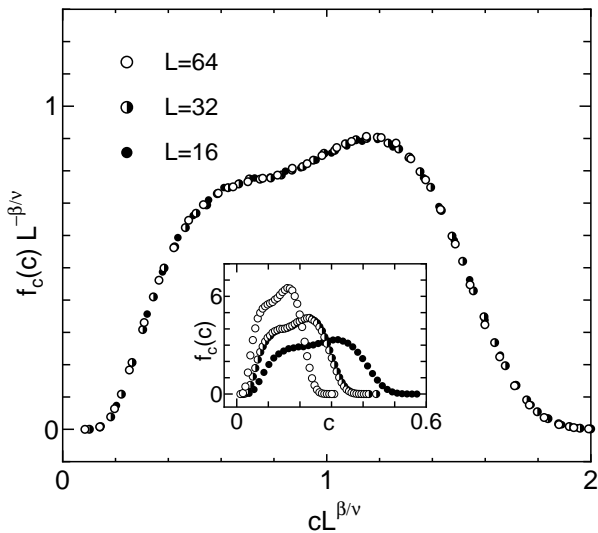}}
\caption{FSS plot of $f_c(c)$ for the 3D bond percolation problem, 
where $\beta/\nu=0.474$. The system sizes are $L=16, 32$, and $64$.
The inset shows the raw data.}
\label{fig_8}
\end{figure}

Finally, we show the FSS plot of the distribution function $f_c(c)$ 
in Fig.~\ref{fig_8}. 
The raw data of $f_c(c)$ for linear sizes $L=16, 32$, and $64$ 
are given in the inset of Fig.~\ref{fig_8}. 
Again, in the percolation problem, the data show very good FSS behavior.

\section{Summary and Discussions}
To summarize, we have given a detailed description of 
the newly proposed PCC algorithm.  We have applied the PCC algorithm 
to the study of the 3D Ising model and the 3D bond percolation problem. 
We have employed a refined analysis of FSS, which uses 
the same scheme as suggested by Ferrenberg and Landau \cite{FerLan}.  
Our results for the Ising model and
the bond percolation problem are consistent with those of 
the previous works \cite{FerLan,LorZif}. 
It is to be noted that we make simulations at a single critical 
point $p_c(L)$ for each system size.  Thus, we need much less 
efforts compared with the usual procedure for making simulations 
at several different temperatures to extract the critical point 
and to estimate critical exponents. 

We have estimated $\nu$ from the temperature derivative of 
the moments.  There is an alternative way to estimate $\nu$ 
without determining $T_c$.  Let us calculate 
the size-dependent $T_c$ in two ways.  We may use different 
criteria for percolating condition; we may use 
different values of $E_p$ which gives $p_c(L)$.  
Then, the difference of two $T_c(L)$'s follows the 
FSS relation as
\begin{equation}
 T_c^{(1)}(L)-T_c^{(2)}(L) = a''L^{-1/\nu}(1 + b''L^{-\omega}), 
\label{tc_diff}
\end{equation}
which also leads to the direct determination of $\nu$.

There are several directions for the application of 
the PCC algorithm.  We can use the PCC algorithm to any problem 
where the mapping to the cluster formalism exists.  
It is straightforward to apply 
this algorithm to the diluted Ising (Potts) models.
The PCC algorithm is quite useful for investigating 
the self-averaging properties of random systems, 
where the distribution of $T_c$ due to randomness is essential. 
It is because we can determine the sample-dependent $p_c(L)$ 
quite easily.  We have already applied the PCC algorithm 
to the 2D site-diluted Ising model \cite{dil2d}, 
and have studied the crossover and self-averaging properties. 
It is also interesting to extend the PCC algorithm 
to the problem of the vector order parameter.  
We have already succeeded in applying the PCC algorithm 
to the classical XY model \cite{XY}, and have shown that 
the PCC algorithm is useful not only for the analysis of 
the second-order transition but also for that of 
the transition of the Kosterlitz-Thouless type. 

In the PCC algorithm, the cluster representation 
is used in two ways.  First, we make a cluster flip 
as in the SW algorithm \cite{SW}. 
Second, we change $p$ depending on the observation 
whether clusters are percolating or not. 
However, the percolating properties are not essential. 
We have used the FSS relation for $E_p$, eq.~(\ref{existence}), 
to determine the critical point. 
We may use quantities other than $E_p$ which have the similar 
FSS relation with a single scaling variable.
We could generalize the PCC algorithm 
for a problem where the mapping to the cluster
formalism does {\it not} exist.  For example, we may study 
the systems with a frustration by the generalized scheme 
of the PCC algorithm.  The application of the PCC algorithm 
to quantum spin systems is also an interesting subject, and 
now in progress.

\section*{Acknowledgments}
We thank N. Kawashima, H. Otsuka, M. Kikuchi,
R. H. Swendsen, and J.-S. Wang for valuable discussions.
The computation in this work has been done using the facilities of
the Supercomputer Center, Institute for Solid State Physics,
University of Tokyo.
This work was supported by a Grant-in-Aid for Scientific Research
from the Japan Society for the Promotion of Science.

\end{document}